\newif\ifarxiv
\arxivtrue  
\RequirePackage{fix-cm}
\documentclass[smallextended,twocolumn]{svjour3}       
\smartqed  
\usepackage{graphicx}
\usepackage{microtype}
\usepackage{amsmath}
\usepackage{amssymb}
\usepackage{mathrsfs}
\usepackage{textgreek}
\spnewtheorem*{theorem*}{Theorem}{\bfseries}{\normalfont}

\ifarxiv
	\renewcommand{\makeheadbox}{}
\else
\fi

\begin{document}
\title{A universal theorem of sensory information\thanks{This manuscript includes work done originally while the author was a professor at the University of Toronto.}
}


\ifarxiv
    \author{Willy Wong
\thanks{Willy Wong \\
            Dept.\ of Informatics, Faculty of Information Science and Electrical Engineering,\\
            Kyushu University.\\
            Tel.: +81-92-802-3805\\
            Fax: +81-92-802-3578\\
            Email: \texttt{willy@inf.kyushu-u.ac.jp}
        }
    }
    \institute{ }
    \date{ }
\else
    \author{Willy Wong}

    \institute{
        Willy Wong \at
        Dept.\ of Informatics, Faculty of Information Science and Electrical Engineering, Kyushu University \\
        Tel.: +81-92-802-3805\\
        Fax: +81-92-802-3578\\
        \email{willy@inf.kyushu-u.ac.jp}
    }
    \date{Received: date / Accepted: date}
\fi
\sloppy

\maketitle

\begin{abstract}
A universal theorem of sensory information, analogous to the second law of thermodynamics, is derived. Beginning from a minimal description of a sensory neuron, a state-space representation of firing rate emerges naturally from Shannon’s measure of information. A special case of this formulation predicts a previously unknown inequality governing sensory adaptation, which was confirmed across different modalities, species, and experimental conditions. Further analysis shows that the firing rate behaves like a state function in thermodynamics, leading to an entropy production equation from which a general law follows: any closed cycle of stimulation yields a non-negative net gain of sensory information.
\end{abstract}

\section{Introduction}
We develop a universal theorem of sensory information --- one that formally mirrors the second law of thermodynamics. While such a claim may seem bold, even implausible within neuroscience, it arises naturally from recent progress in the modelling of sensory neurons which is supported by experimental evidence across multiple modalities and species. Careful analysis of this body of work reveals a set of generalizable principles from which a generic and axiomatic description of sensory information can be obtained, culminating in an inequality that appears to govern information acquisition in many, if not all, living systems.

To appreciate the significance of this result, we must first recognize the conceptual challenges inherent in modelling complex systems. Developing models in neuroscience is far from straightforward. Bottom-up, physics-based approaches often lack sufficient constraints on their many parameters e.g. \cite{almog2016realistic}, while black-box models offer no recourse for understanding or justifying the values its parameters can take \cite{herz2006modeling}. As von Neumann once stated, a model with sufficient number of free parameters can always be made to fit any arbitrary dataset \cite{dyson2004meeting}. In this context, the most stringent test of a theory lies in its ability to predict and anticipate phenomena that have yet to be observed. 

A further challenge arises from the intrinsic variability and noise of biological systems, which differ markedly from the controlled systems of physics or chemistry. Variability and diversity are often fundamental in biology. Replication across different conditions is therefore essential to establish whether a finding is universal or restricted to a single case. The need for convergence is captured by Whewell’s principle of consilience: independent lines of evidence must align before a claim can be considered true \cite{whewell1840philosophy,wilson1999consilience}. Together, these considerations demonstrate that establishing universality in biology requires support across a diverse set of conditions including testing in different species, experimental conditions, and even different laboratories.

When consistent patterns do arise, there remains the question of formalization: can these patterns be used to formulate mathematical theorems? Theorems represent the highest form of certainty and rigour, yet unlike in mathematics or physics, developing theorems in biology is challenging because, as discussed earlier, the observations used to form the axioms are often context-dependent and sensitive to conditions. In essence, the challenge is to identify principles that are truly invariant and remain valid regardless of biological context.

This work concerns the application of information theory to the study of sensory systems. Neuroscience has long used information theory to analyze spike trains, probing the entropy and efficiency of coding. Some early work introduced the notion of the efficient coding hypothesis \cite{attneave1954some,barlow1961possible}, which was later expanded to show that single neurons can approach theoretical limits of information transmission \cite{strong1998entropy,borst1999information}. Yet despite their wide adoption, these studies should be viewed with some caution. They rest on assumptions about the nature of the neural code and the methods used to estimate it. As such these calculations of information remain open to interpretation. In contrast, Norwich \cite{norwich1977information,norwich1993information} and Wong \cite{wong2020rate} have proposed an alternative approach that grounds information in measurable physiological variables rather than coding efficiency. The present work builds upon and extends this line of reasoning.

Building on that foundation, we introduce the concept of the \textit{ideal sensory unit}: a minimal and general mathematical representation that accounts the essential mechanism of sensory transduction in its most reduced form. A special case of this framework recovers a previously published theory of sensory processing and its prediction, which together satisfy the stringent validation criteria discussed earlier. On this basis, we derive a fundamental theorem governing the acquisition of sensory information. In doing so, the work moves beyond descriptive or phenomenological models toward a principle-based understanding of sensory processing.

\section{The ideal sensory unit}
The concept of the ideal sensory unit captures sensory processing at its fundamental core --- a universal mechanism independent of biological implementation, sensory modality, or species. Here, the input to the unit is sensory intensity $\mu(t)$ and the output is the firing rate response $F(t)$ as measured at the primary afferent or first-order neuron through the number of spikes generated per second.  From a continuous time system perspective, the ideal sensory unit can be modelled as a nonlinear, time-invariant system which is causal with memory (i.e. depends only on past and present inputs).  We also take the response of the system to a step function—corresponding to a sudden onset of stimulation—to be discontinuous.  That is, the firing rate exhibits an instantaneous jump at stimulus onset.  Written in terms of a state-space representation we have
\begin{align}
\label{statespacerepresentation1}
\dot{m}(t) &=  g\left( m(t), \mu(t)\right), \\
 F(t) &=  f\left(m(t),\mu(t)\right)
\end{align}
where $f$ and $g$ are unknown and possibly non-linear functions.  Here, $F$ is the output variable, $\mu$ the input and $m$ the state variable.  The task is to reveal the form of $f$ and $g$, and elucidate the role of $m$ in the sensory process.  For convenience, a summary of the principal variables and parameters used in this paper is provided in Table 1.
%
\subsection{Connection to Shannon's information}
In physics, the mathematical form of a law often reflects symmetries and invariances in the natural world: Newton’s equations, for example, are invariant under time reversal. If the equations governing the sensory response are to reflect the perceptual process, they too should embody the structural properties of perception itself.

A natural question, then, is how to distill the process of perception. One approach is to think of perception as the act of reducing uncertainty. In one view, the perceptual system continuously infers the most likely state of the world from sensory data \cite{von1867handbuch,gregory1980perceptions,clark2013whatever}.  Alternatively, perception may be described as undergoing an \textit{intellego process}—the act of selecting one outcome from among several possible alternatives \cite{norwich1991fundamental}. To perceive that a hat is red, for example, is to select “red” from among the possible colours it might have taken. In this sense, perception can be thought of as involving inference and selection, linking it fundamentally to the concepts of uncertainty and information.

Following Shannon \cite{shannon1948mathematical}, Faddeev showed that a measure of uncertainty $H$ can be derived uniquely from three simple assumptions. Let $H(p_1, p_2, \dots, p_n)$ be a real-valued function defined for all probability distributions $(p_1, p_2, \dots, p_n)$ such that $p_i \ge 0$ and $\sum_{i=1}^n p_i = 1$. If $H$ satisfies the following conditions:
\begin{enumerate}
\item Continuity: $H(p_1, p_2, \dots, p_n)$ is a continuous function of $(p_1, p_2, \dots, p_n)$.
\item Symmetry: \\ $H(p_1, p_2, \dots, p_n) = H(p_{\pi(1)}, p_{\pi(2)}, \dots, p_{\pi(n)})$ for any permutation $\pi$ of $\{1,2,\dots,n\}$.
\item Recursivity: If a probability $p_n$ is divided into two parts $q$ and $r$ such that $q+r=p_n$, then
\begin{align*}
H(p_1, p_2, \dots, p_{n-1}, q, r)
  &= H(p_1, p_2, \dots, p_{n-1}, p_n) \\
  &\quad +\, p_n\, H\!\left(\frac{q}{p_n}, \frac{r}{p_n}\right).
\end{align*}
\end{enumerate}
Then it follows that
\begin{equation}
\label{shannon}
H(p_1, p_2, \dots, p_n) = -K \sum_{i=1}^n p_i \log p_i,
\end{equation}
for some positive constant $K>0$ \cite{faddeev1956concept}.

Remarkably, these assumptions appear to mirror the very structure of perception. Continuity ensures perceptual stability—small changes in the stimulus should not cause large perceptual changes; symmetry reflects perceptual objectivity—labelling should not alter perception; and recursivity corresponds to perceptual resolution—an organism is more perceptive when finer categories can be resolved. Thus, the mathematical form of $H$ appears to embody the same logical requirements that govern perception itself.

In the case of intensity perception, firing rate is the principal driver of sensory experience. Across sensory modalities, firing rate increases monotonically with stimulus intensity, and empirical scaling laws show a logarithmic dependence that mirrors Shannon’s form of uncertainty. If both firing rate and uncertainty reduction vary monotonically with stimulus intensity, their functional dependence must share the same qualitative form. These considerations motivate the proportional relation
\begin{equation}
\label{f=kh}
F = kH,
\end{equation}
where $k$ is a proportionality constant \cite{norwich1977information,norwich1993information}. Here $F$ represents the physiological correlate of uncertainty, while $H$ provides its mathematical form. The proportionality $F = kH$ therefore provides a testable bridge between the measurable neural response and the abstract quantity of information or uncertainty.  

Having established this connection conceptually, we next examine how this relationship arises mechanistically within the ideal sensory unit.

\subsection{Sensation as a measurement process}
At the level of the ideal sensory unit, the process of “selecting one alternative from many” corresponds to sampling a noisy signal to estimate its mean value. Sampling is assumed to be carried out in real time over independent epochs, ensuring that each epoch provides an unbiased update of the estimated mean. When a signal with mean $\mu$ and variance $\sigma^2$ is sampled repeatedly, provided that the number of samples drawn is large enough, the resulting uncertainty in the mean will be normally distributed by the central limit theorem with identical value of mean and variance $\sigma^2/m$, where $m$ is the number of samples. Moreover, in any measurement process there is typically also representational noise, i.e., the noise arising from the representation or storage of the measured value (e.g. resolution error). If we think of this error as being normally distributed with zero mean and variance $\sigma_R^2$, and that measurement error is independent of representation error, the entire continuous Shannon uncertainty or differential entropy can be calculated to be
\begin{equation}
\label{hequation}
H=\tfrac{1}{2}\log(\sigma_R^2+\sigma^2/m)+\text{constant}
\end{equation}
where the additive constant contains both a numerical constant together with the dependence on measurement units.

Three final observations are required before we are complete.  First, greater variability is generally associated with signals of greater magnitude --- a common pattern observed in nature and data where variance scales with the mean $\sigma^2(\mu)$, often referred to as fluctuation scaling (e.g. \cite{eisler2008fluctuation}).  Second, the sample size tends toward a unique optimal value $m_{eq}$ in a steady, one-way manner. Once it reaches this optimal value, the sample size no longer changes.  Finally, since variance scales with mean, it stands to reason that the optimal sample size must also scale with mean, i.e. $m_{eq}(\mu)$, although not necessarily in the same manner.

Summarizing, the behaviour of the ideal sensory unit can be characterized by the following state-space representation:
\begin{align}
\dot{m} &= g\left(m, m_{eq}(\mu)\right), \label{diffeqn} \\
H &= \tfrac{1}{2}\log\!\left(\sigma_R^2 + \frac{\sigma^2(\mu)}{m}\right) + \text{constant}, \label{prehequation}
\end{align}
where the firing rate is given by $F = kH$. This formulation is subject to three additional constraints:  (1) $\sigma_R^2$ is constant;  (2) $m$ tracks $m_{eq}(\mu)$ monotonically, with $g(m, m_{eq}) = 0$ if and only if $m = m_{eq}$; and  (3) $\sigma^2$ and $m_{eq}$ are both continuous, non-decreasing functions of $\mu(t)$.  Although $m$ represents sample size, we treat it as a continuous variable—an approximation justified in the large $m$ limit. In passing, we note that this characterization amounts to calculating the uncertainty associated with a generic measurement process (repeated sampling and averaging) and calculating the firing rate with that uncertainty. If correct, this model satisfies the requirement of being independent of biological implementation or sensory modality. At its most fundamental level, the sensory system is about information processing, and the framework of the ideal sensory unit places information at the centre of its mathematical representation.

\subsection{Example}
\label{example}
Next, we show how the ideal sensory unit reproduces key results and predictions reported previously.

A standard way to analyze a nonlinear state-space system is to linearize it around a fixed point. In this case, the fixed point occurs when the sample size reaches its steady-state or equilibrium value $m_{eq}$. For constant stimuli, expanding $\dot{m} = g(m, m_{eq})$ to first order about $m = m_{eq}$, and noting that $g(m_{eq},m_{eq})=0$ (sampling ceases at equilibrium), yields
\begin{equation}
\label{relax}
\frac{dm}{dt} = -a (m - m_{eq})
\end{equation}
where $a = -(dg/dm)\big|_{m = m_{eq}} > 0$ is a positive constant reflecting relaxation toward $m_{eq}$. Moreover, if the distribution of the input or stimulus signal follows the Tweedie family of distributions which provides a general form linking variance and mean for a wide class of natural signals, we have $\sigma^2 \propto \mu^p$ and $m_{eq} \propto \mu^{p/2}$. Finally, taking the constant in (\ref{hequation}) to be equal to $\tfrac{1}{2}\log(\sigma_R^2)$, we obtain
\begin{gather}
\label{net}
H = \frac{1}{2} \log \left( 1 + \frac{\beta (I + \delta I)^p}{m} \right) \\
m_{eq} = (I + \delta I)^{p/2}
\end{gather}
where the input $\mu=I+\delta I$ is taken to be stimulus intensity $I$ plus additive noise $\delta I$. The five parameters, $k, \beta, p, \delta I, a$ are assumed to be positive, and help ensure the equation is dimensionally consistent.

These are the same equations that have evolved across a series of publications beginning in the 1970's \cite{norwich1977information,norwich1993information,norwich1995universal,wong1997physics,wong2020rate}, and have been shown to reproduce experimental observations across a wide range of conditions. We can use these equations to compute the time-dependent firing rate $F(t)$ for any given stimulus $I(t)$.  The parameters are estimated through data fitting, although in practice their estimation can be challenging due to overfitting.

Adaptation refers to the property whereby a neuronal response to a stimulus rises from its initial spontaneous value, reaches a peak shortly after stimulation, and then falls monotonically to a steady-state value that persists until the stimulus ceases. When the stimulus is held constant for $t \ge 0$, $m_{eq}$ is constant, and the solution to (\ref{relax}) is
\begin{equation}
\label{msolution}
m(t) = m(0)e^{-at} + m_{eq} (1 - e^{-at})
\end{equation}
By continuity, $m(0)$ equals to the equilibrium sample size when $I = 0$, i.e. $m(0) = \delta I^{p/2}$ \cite{wong2020rate}. Substituting (\ref{msolution}) into (\ref{net}) and (\ref{f=kh}) gives the full time-dependent firing-rate response during adaptation:
\begin{equation}
\label{adaptation}
F(I, t) = \frac{k}{2} \log \left( 1 + \frac{\beta (I + \delta I)^p}{ \delta I^{p/2} e^{-at} + (I + \delta I)^{p/2} (1 - e^{-at}) } \right)
\end{equation}
From this, we identify the three fixed points of adaptation --- the spontaneous rate $\text{SR}=F(0,\infty)$, the peak rate $\text{PR}=F(I,0)$, and the steady-state rate $\text{SS}=F(I,\infty)$:
\begin{gather}
\label{SRPRSS}
\text{SR} = \frac{k}{2} \log \left( 1 + \beta \delta I^{p/2} \right) \\
\text{PR} = \frac{k}{2} \log \left( 1 + \frac{\beta (I + \delta I)^p}{ \delta I^{p/2} } \right) \\
\text{SS} = \frac{k}{2} \log \left( 1 + \beta (I + \delta I)^{p/2} \right)
\end{gather}

These quantities, together with the curvature of the logarithmic function, lead to a previously undiscovered property of sensory adaptation. The following elegant inequality connects the three points of the adaptation curve:
\begin{equation}
\label{inequality}
\sqrt{\text{PR} \times \text{SR}} \le \text{SS} \le \left(\text{PR} + \text{SR}\right)/2
\end{equation}
That is, the steady-state activity of a sensory neuron must always be bounded by the geometric and arithmetic means of its spontaneous and peak responses \cite{wong2023fundamental}. This result reveals a remarkable mathematical simplicity underlying sensory adaptation. The simplicity, however, belies the depth of its predictive reach: a single inequality derived from first principles reproduces adaptation behaviour in a variety of modalities, despite differences in receptor mechanisms and method of stimulation.


%
The inequality was evaluated across two comprehensive analyses encompassing 40 separate studies and more than 400 individual recordings of adaptation from different sensory modalities, animal species, and laboratories. The first analysis examined all available auditory adaptation data from the 1970's to present. Hearing is particularly well suited for repeated adaptation experiments because auditory stimuli can be precisely controlled and single-unit responses readily isolated. A number of studies provide adaptation curves across a wide range of stimulus intensities, enabling direct comparison of observed steady-state firing rates against the theoretical bounds. In each case, the steady-state firing rate (SS) was plotted against the corresponding peak response (PR) across stimulus levels and compared with the predicted geometric and arithmetic mean limits of (\ref{inequality}). \textit{No parameter fitting was required}.  The only parameter, spontaneous activity (SR), was obtained directly from the recordings. Fig. 1, reproduced from \cite{wong2023fundamental}, shows empirical firing-rate trajectories superimposed on the theoretical bounds of (\ref{inequality}) with very few outliers, demonstrating that steady-state responses lie reliably within—and often approach—the predicted arithmetic and geometric mean limits.

\begin{figure*}[!h]
\begin{center}
\includegraphics[width=.89\textwidth]{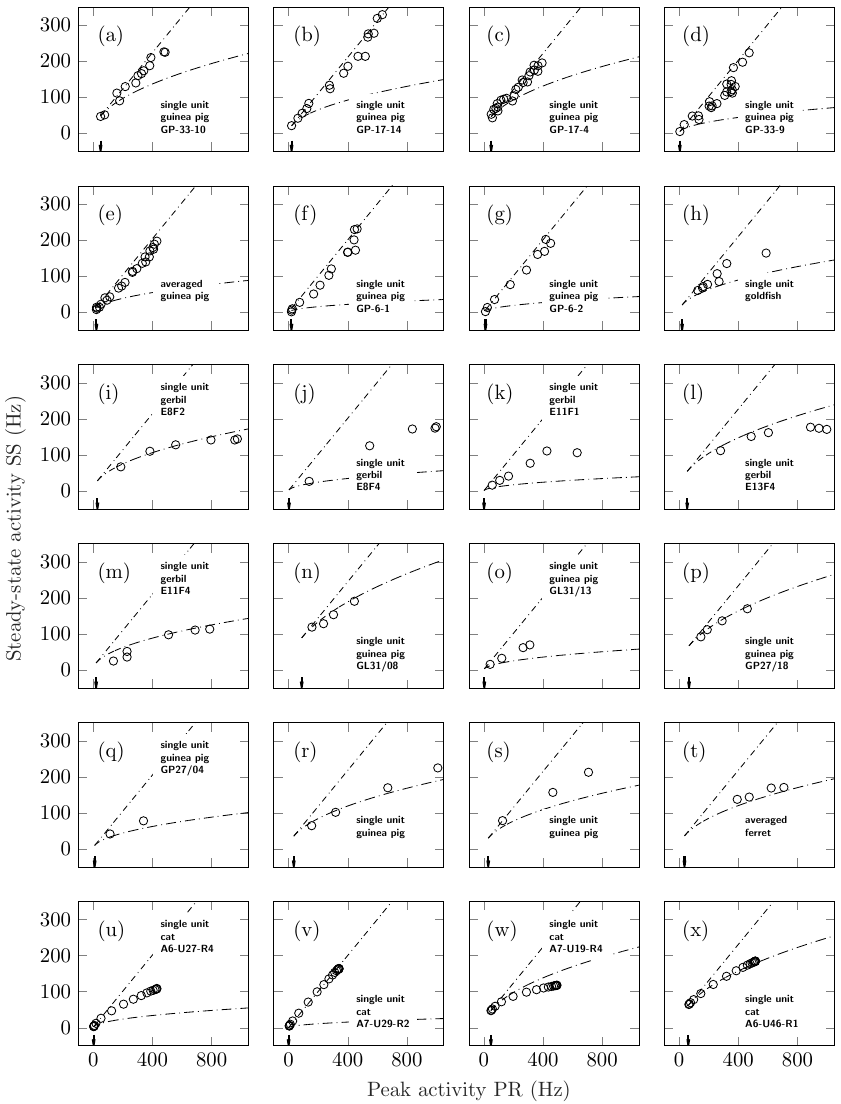}
\caption{Figure reproduced from \cite{wong2023fundamental}. Steady-state activity plotted versus peak activity for a number of auditory studies.  In all panels, the dashed lines show the theoretical upper and lower bounds of (\ref{inequality}).  No fitted parameters were required to plot these bounds.  SS vs PR from (a-g) single or averaged guinea pig fibre recordings (figs. 11a-d, 12, 17a and 4a from \cite{smith1975short}); (h) saccular nerve fibres of goldfish (fig. 3 from \cite{fay1978coding}); (i-m) single fibre gerbil recordings (figs. 4 and 5 from \cite{westerman1984rapid}); (n-q) single guinea pig fibre recordings (figs. 1 and 2 from \cite{yates1985very}); (r-s) single guinea pig fibre recordings (figs 3a and 3b from \cite{muller1991relationship}); (t) averaged ferret data (fig. 6 from \cite{sumner2012auditory}); (u-x) single cat fibre recordings (figs. 12e-h from \cite{peterson2021simplified}).  The spontaneous activity of each unit is indicated by an arrow pointing towards the x-axis.}\label{fig1}
\end{center}
\end{figure*}

In a second comprehensive study, the analysis was extended to other sensory modalities \cite{wong2021consilience}. Since spontaneous activity is not always reported, a different approach was taken. When the steady-state response is approximated by the lower bound of the inequality, $\text{SS} \approx \sqrt{\text{PR} \times \text{SR}}$, a double-logarithmic plot of SS versus PR yields a straight line with slope 1/2. In a survey across eight major sensory modalities (proprioception, touch, taste, hearing, vision, smell, electroreception, and temperature) and four major animal phyla (Chordata, Arthropoda, Mollusca, and Cnidaria), most datasets exhibited slope near 0.5 with high correlation ($R^2 > 0.9$). Taken together, these findings constitute one of the most extensive cross-modal comparisons of sensory adaptation to date.

The convergence of findings here implies a certain degree of \textit{universality}. Neurons, of course, do not operate in identical mechanistic fashion, yet empirically we observe the same quantitative relation holding across a diverse set of modalities, species, and  measurement paradigms. Remarkably, data recorded nearly a century ago by pioneers like Adrian and Zottermann \cite{adrian1926impulsesb} show the same
adherence to the inequality as modern recordings, underscoring the robustness of (\ref{inequality}) across time and methodology. This coherence exemplifies the principle of consilience: independent lines of evidence converging on a common truth.

What do these results imply for the mechanics of the model? The primary part of (\ref{prehequation}) is the term $\sigma^2(\mu)/m$. The peak response shows that $\sigma^2$ increases instantaneously at stimulus onset, while $m$ evolves continuously across the boundary. Moreover, since peak firing increases with stimulus intensity, this implies that $\sigma^2$ must be a monotonically increasing function of the input. Steady-state activity also rises with intensity, albeit more slowly, suggesting that given the ratio $\sigma^2(\mu)/m_{eq}$, the equilibrium sample size $m_{eq}$ must likewise increase with stimulation level. However, not all aspects of the theory were tested empirically including, in particular, the time course of adaptation. This motivates returning to the broader framework of the ideal sensory unit, which imposes weaker conditions than the model considered here. For example, since $g$ in (\ref{diffeqn}) was not assumed to be analytic, the ideal sensory unit accommodates a wider range of relaxation behaviours, including power-law adaptation \cite{drew2006models}.

The generality of the approach suggests that the ideal sensory unit may extend beyond a purely biological interpretation. In fact, the structure of the equations hints at similarities to theories in physics, a connection that will be explored next.

\section{Correspondence with thermal physics}
We now extend the framework by establishing an analogy between sensory processing and thermal physics. The equations describing the ideal sensory unit bear a striking resemblance to those of statistical physics. $H$ is computed via a Shannon–Boltzmann entropy, with $H$ related to a measurable quantity (firing rate) through the relation $F = kH$. Moreover, using a well-known result from information theory \cite{shannon1948mathematical}, we can write for the sample mean $\theta$,
\begin{equation}
\label{conditioning}
H(\theta \mid X_1, \ldots, X_m) \le H(\theta)
\end{equation}
where $X_1, \ldots, X_m$ are samples drawn by the unit, $H(\theta)$ the prior entropy, and $H(\theta \mid X_1, \ldots, X_m)$ the posterior entropy. That is, $H$ must remain the same or decrease as additional samples or measurements are taken. Additionally, since $dm/dt \ge 0$ and $F = kH$, we conclude that $dF/dt \le 0$. At first glance, this suggests that firing rate must decrease over time—which is indeed what happens during adaptation—but this interpretation is not entirely correct: in general, $m$ can both increase or decrease, and therefore $dF/dt$ can take either sign. Nevertheless, we note a strong resemblance to Boltzmann’s equations:
\begin{gather}
\label{Boltzmann}
S = k_B H, \\
H = -\!\int p(x)\log p(x)\,dx, \\
\frac{dS}{dt} \ge 0.
\end{gather}

The variables on both sides, of course, describe different domains: the former quantifies uncertainty in a sensory measurement, the latter quantifies uncertainty in phase space. Nevertheless, both share a common mathematical structure—the logarithmic measure of variability—and both appear to obey an inequality capturing directionality in their respective domains. This analogy therefore suggest that information processing in sensory systems can be expressed through equations that mirror the form—though not necessarily the substance—of thermodynamic laws. 

\subsection{Firing rate response as a state function}
Before proceeding further, it is important to clarify how the term \textit{state} is used in both control theory and in thermodynamics. In control theory, a state variable summarizes the entire history of the system up to a given time; in our formulation, the variable 
$m$ plays this role. In thermodynamics, a state refers to an equilibrium condition that is fully described by its state variables (e.g., $P$,$V$ and $T$). Here, the external input $\mu$ serves as the thermodynamic analogue of a state variable, since $m_{eq}(\mu)$ specifies the corresponding equilibrium value of $m$.

Moreover, a \textit{state function} is a property whose value depends only on the current equilibrium state of the system, independent of the path or process used to reach that state. An example of a state function is thermodynamic entropy, for which the change in entropy is only dependent on the initial and final states, and not the path by which it gets there. What about the sensory domain.  At equilibrium $\dot{m} = 0$, and the system settles to a unique value $m=m_{eq}$. Since both $\sigma^2$ and $m_{\text{eq}}$ depend on $\mu$, the equilibrium sensory entropy $H_{\text{eq}} = \tfrac{1}{2} \log \left( \sigma_R^2 + \sigma^2/m_{eq} \right) + \text{constant}$ is purely a function of $\mu$. Consequently, the equilibrium value of $H$ depends only on the current input $\mu$ and not on its past history; hence both $H$ and the firing rate $F=kH$ are state functions.

Is this in fact supported by experiment? While this is clearly an idealization of how real neurons function, experiments suggest that the response typically settles into a steady-state level that depends only on $\mu$ and is independent of the path taken to reach it, even though  neurons can have difficulty maintaining sustained activity over very long periods of time (minutes or longer) (e.g. \cite{javel1996long}). Some of the earliest experiments testing this hypothesis were conducted in the pioneering studies of Adrian and Zottermann \cite{adrian1926impulsesc}. Fig. 2 shows the firing rate in proprioceptive neurons measured under varying slopes of increasing loads (ramps). Despite differences in ramp speed, the neurons ultimately reached the same steady-state firing rate. Subsequent studies have reported similar findings (e.g., \cite{boyd1953proprioceptive}).

While most neuroscientists tacitly assume that neural activity behaves like a state function (e.g. the firing rate returns to baseline after stimulation), what is often missing is a formal proof that an equilibrium value exists, is unique, and is independent of the path taken to reach it. Here, we have shown that $F$ behaves as a state function within a dynamical, information-theoretic framework, a result that holds robustly across species, sensory modalities, and experimental conditions. Next, this observation is used to investigate path dependency in calculating changes in firing rates.

\begin{figure}[!h]
\begin{center}
\includegraphics[width=.49\textwidth]{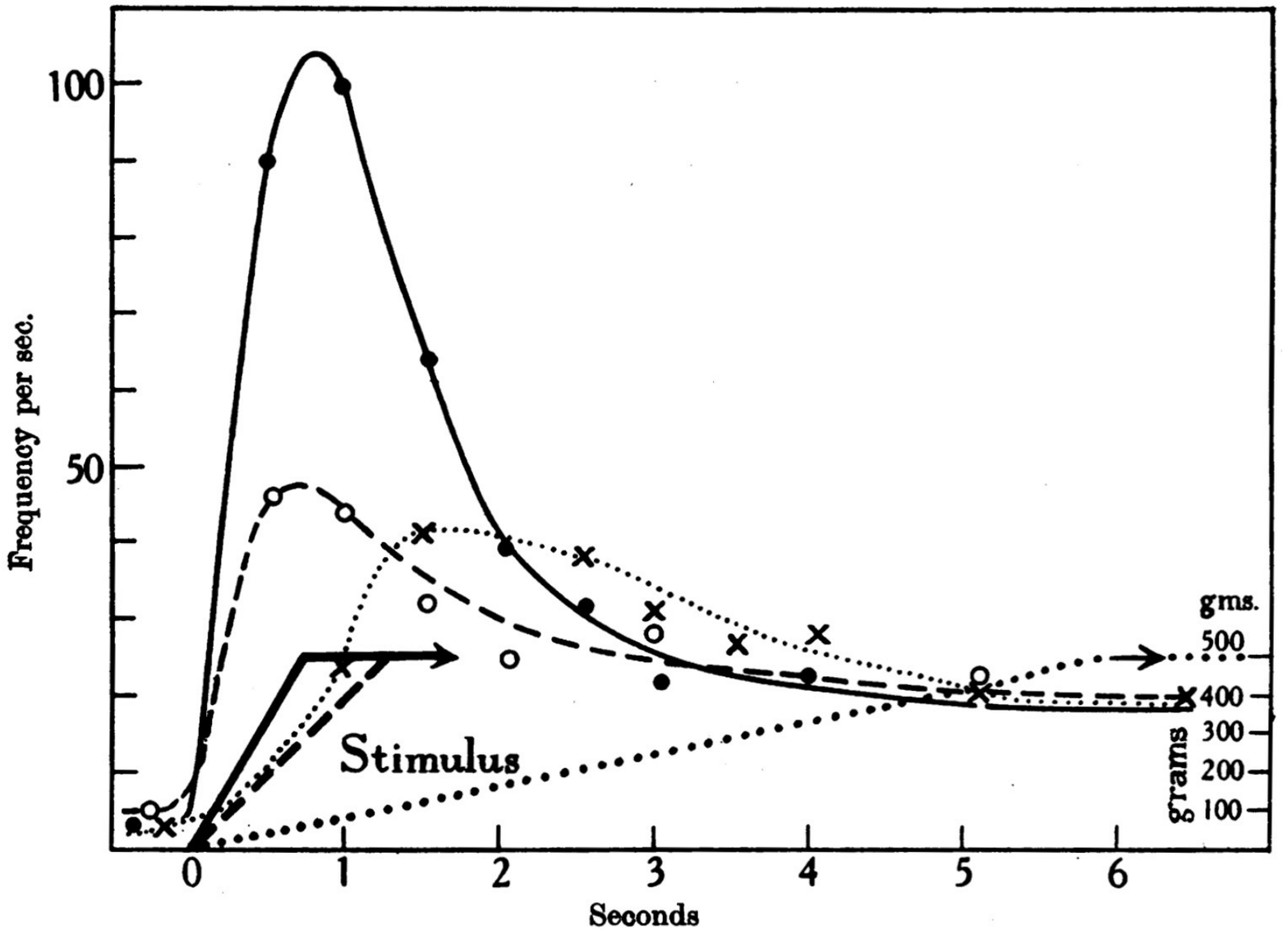}
\caption{Figure reproduced from \cite{adrian1926impulsesc}, illustrating that the final equilibrium value of the firing rate does not depend on the rate at which the load is increased on the mechanoreceptors of the cat. The solid, dashed, and dotted lines represent three different rates of ramp increase, along with their associated responses.}\label{fig2}
\end{center}
\end{figure}

\subsection{Entropy balance equation}
In classical thermodynamics, the entropy balance equation expresses the rate of change of entropy as the sum of an entropy flux term and an entropy production term:
\begin{equation}
\label{thermodynamicentropybalance}
dS = \frac{\delta Q}{T} + \delta S_{\text{prod}},
\end{equation}
where $Q$ is the exchanged heat, $T$ the temperature, and $S_{\text{prod}}$ the entropy produced internally. The symbol $d$ denotes an exact differential, while $\delta$ denotes an inexact one that depends on the path of integration. For a cyclic process in which the system returns to its initial state, the integral of the exact differential $dS$ is zero:
\begin{align}
\label{clausius}
\oint dS &= \oint \frac{\delta Q}{T} + \oint \delta S_{\text{prod}} = 0.
\end{align}
From Clausius’ inequality, $\oint \delta Q/T \le 0$, and the second law of thermodynamics follows immediately:
\begin{equation}
\label{thermodynamicsecondlaw}
\oint \delta S_{\text{prod}} \ge 0.
\end{equation}
Any irreversible process results in a net positive production of entropy, transferring entropy to the surroundings.

By analogy, for the ideal sensory unit—where both $F$ and $H$ are state functions—we can formulate an equivalent balance equation:
\begin{align}
\label{sensoryentropybalance}
dH &= \frac{\partial H}{\partial \mu}\,d\mu
+ \frac{\partial H}{\partial m}\,dm \\
&= \delta H_{\text{flux}} + \delta H_{\text{relax}}.
\end{align}
We identify a flux term $\delta H_{\text{flux}}$ associated with changes in the external stimulus $\mu$, and a relaxation term $\delta H_{\text{relax}}$ arising from changes due to sampling. Their sum forms an exact differential, yet each term individually is path dependent because $\mu$ and $m$ are \textit{interdependent}; hence $H$ cannot be written in an additively separable form of two variables. The path dependence of $\delta H_{\text{relax}}$ will be illustrated by an example next.

\subsection{High/low stimulus thought experiment}
\label{highlow}
Consider the ideal sensory unit already at steady state under input $\mu_1$. The stimulus is then abruptly increased to $\mu_2 \ge \mu_1$ and held there steadily before returning to $\mu_1$,  where it remains until the response again reaches steady state. This represents the simplest perceptual cycle: a transient increase in stimulus followed by the recovery to baseline. Since $m$ evolves continuously while $\sigma^2$ changes abruptly during the transition $\mu_1 \!\rightarrow\! \mu_2$, the response overshoots at onset and then relaxes monotonically as $m$ tracks $m_{eq}(\mu_2)$. When the stimulus returns to $\mu_1$, the reverse occurs, producing an undershoot before the system settles back to its initial state.

Let us denote $\sigma_1^2=\sigma^2(\mu_1)$ and $\sigma_2^2=\sigma^2(\mu_2)$.  Moreover, $m_1=m_{eq}(\mu_1)$, $m_2=m_{eq}(\mu_2)$, and $m'$ is the sample size when the stimulus is returned back to $\mu_1$ with $m_1 \le m' \le m_2$.  The net  contribution to $H_{\text{relax}}$ over this cycle can be obtained by summing the difference in $H$ during the two adaptation periods to obtain:
\begin{align}
\label{thoughtexperiment}
\Delta H_{\text{relax}}
  &= H(\mu_2,m')-H(\mu_2,m_1) \notag \\
  &\qquad
	+H(\mu_1,m_1)-H(\mu_1,m')  \\
  &= \frac{1}{2}
     \log\!\left(
       \frac{\sigma_R^2+\sigma_2^2/m'}{\sigma_R^2+\sigma_2^2/m_1}
     \right) \notag \\
  &\qquad
     + \frac{1}{2}
       \log\!\left(
         \frac{\sigma_R^2+\sigma_1^2/m_1}{\sigma_R^2+\sigma_1^2/m'}
       \right) \\
  &\le 0.
\end{align}
where the last inequality holds whenever $\sigma_2^2 \ge \sigma_1^2$ and $m' \ge m_1$. Thus, over a complete stimulus cycle, the relaxation component of entropy exhibits a net decrease in relaxation entropy. This asymmetry arises because the change in uncertainty during the upward transition differs from that during the downward transition.  For the same sampling ``effort'', uncertainty reduction is more prominent when the value of variance is larger.

This result motivates the next development, which examines the behaviour of relaxation entropy over an arbitrary stimulus cycle.

\section{An information inequality}

As $H$ is a state function, its closed-loop integral must vanish, analogous to entropy in thermodynamic cycles:
\begin{equation}
\label{statefunction}
\oint dH = 0.
\end{equation}
However, when we decompose $dH$ into its flux and relaxation components, the relaxation term behaves differently from $H$. We now define sensory information, $\mathscr{I}$, as the negative of the relaxation component of $H$:
\begin{equation}
\label{sensoryinformation}
\mathscr{I} = -\,\delta H_{\text{relax}}.
\end{equation}
That is, for a fixed input $\mu$, the sensory information represents the change in uncertainty resulting from additional sampling of the stimulus.


\begin{theorem*}[Sensory information inequality]
Let $\mu(t)$ be a cyclic stimulus that generates a trajectory $(\mu,m)$ in state space. The change in sensory information over this cycle obeys a second law-like inequality:
\begin{equation}
\label{sensorysecondlaw}
\oint_C d \mathscr{I} \ge 0.
\end{equation}
\end{theorem*}

Two proofs are provided. The first is geometric, relying on the curvature of $H(\mu,m)$; the second is motivated by the types of stimuli used when conducting adaptation experiments. 

\begin{proof}[1]
Let $A$ be a closed, bounded region in $(\mu, m)$ space with a piecewise-smooth, simple, counter-clockwise boundary $C$. By Green's theorem, the net change in sensory information over $C$ can be written as
\begin{equation}
\label{green}
\oint_C d\mathscr{I}
= -\oint_C \frac{\partial H}{\partial m}\,dm
= -\iint_A \frac{\partial^2 H}{\partial m\,\partial \mu}\,dA,
\end{equation}
The mixed partial derivative is evaluated to be
\begin{equation}
\label{mixedderivative}
\frac{\partial^2 H}{\partial m\,\partial \mu}
= -\frac{\sigma_R^2}{2m^2\left( \sigma_R^2 + \sigma^2/m \right)^2} \cdot \frac{d\sigma^2}{d\mu},
\end{equation}
which is $\le 0$ in $A$ provided that $\sigma^2(\mu)$ is non-decreasing. Since the integrand is non-positive everywhere, and thus the area integral is also non-negative, therefore $\oint_C d\mathscr{I} \ge 0$ as required.

Physically, however, not every loop in $(\mu, m)$ space corresponds to a realizable stimulus--response cycle. Suppose we begin at point $(\mu_0, m_0)$ where $m_0$ is the equilibrium sample size for $\mu_0$. As the input $\mu$ changes along the loop, the system dynamics dictate that $m$ must evolve toward the corresponding equilibrium value $m_{eq}(\mu)$. Therefore, to ensure that the path can traverse toward higher $m$ as $\mu$ increases---and likewise toward lower $m$ as $\mu$ decreases---it is necessary that $m_{eq}(\mu)$ be a non-decreasing function.  In summary, while the mathematical inequality holds under the condition that $\sigma^2(\mu)$ is non-decreasing, physical realizability of the loop further requires that $m_{eq}(\mu)$ also be non-decreasing.
\end{proof}

\begin{proof}[2]
This proof evaluates $\mathscr{I}$ for discrete, pedestal-like changes in $\mu$, then passes to the continuum limit. Consider an ideal sensory unit fully adapted to $\mu_0$, then sequentially driven through values $\mu_1, \mu_2, \dots, \mu_{N-1}$ before returning to $\mu_0$. Associated with these levels are variances $\sigma_0^2,\sigma_1^2,\dots,\sigma_{N-1}^2$ and equilibrium sample sizes $m_0,m_1,\dots,m_{N-1}$. Let $m'_j$ denote the sample size immediately before transition from $\mu_j$ to $\mu_{j+1}$, with $m'_0=\!m_0$.

The total change in information $\Delta \mathscr{I}_N$ over $N$ inputs is
\begin{equation}
\label{delta i}
\begin{split}
\Delta \mathscr{I}_N
  &= \tfrac{1}{2}\log\!\left(
       \frac{\sigma_R^2+\sigma_1^2/m_0}{\sigma_R^2+\sigma_1^2/m'_1}
     \right)
   + \tfrac{1}{2}\log\!\left(
       \frac{\sigma_R^2+\sigma_2^2/m'_1}{\sigma_R^2+\sigma_2^2/m_2}
     \right) \\
  &\quad
   +\dots+ \tfrac{1}{2}\log\!\left(
       \frac{\sigma_R^2+\sigma_{N-1}^2/m'_{N-2}}{\sigma_R^2+\sigma_{N-1}^2/m'_{N-1}}
     \right) \\
  &\quad
   + \tfrac{1}{2}\log\!\left(
       \frac{\sigma_R^2+\sigma_0^2/m'_{N-1}}{\sigma_R^2+\sigma_0^2/m_0}
     \right).
\end{split}
\end{equation}
Exponentiating both sides and defining $x_j=m'_j$ and $y_j=\sigma_j^2/\sigma_R^2$, we obtain
\begin{align}
\label{fraction}
&e^{2\Delta \mathscr{I}_N} = \notag \\
&\hspace{2em} \frac{(x_0+y_1)(x_1+y_2)\dots(x_{N-2}+y_{N-1})(x_{N-1}+y_0)}
{(x_1+y_1)(x_2+y_2)\dots(x_{N-1}+y_{N-1})(x_0+y_0)},
\end{align}
which must be shown to be $\ge 1$.

Before addressing the general case, it is instructive to consider two special cases often observed experimentally (e.g. \cite{schafer1988analysis}). (i) For monotonic staircases ($\mu_0 \le \mu_1 \le \dots \le \mu_{N-1}$), both $y_j$ and $x_j$ are ordered sequences, and by a corollary of the rearrangement inequality \cite{hardy1952inequalities,oppenheim1954inequalities},
\begin{equation}
\label{reverse}
\prod_{j=0}^{N-1}(a_j+b_j)
\le \prod_{j=0}^{N-1}(a_j+b_{\sigma(j)})
\le \prod_{j=0}^{N-1}(a_j+b_{N-1-j}),
\end{equation}
the denominator in (\ref{fraction}) attains the minimum value, yielding $e^{2\Delta \mathscr{I}_N}\ge1$. (ii) When each input transition is held long enough for $m'_j=m_j$, the same conclusion holds because even though $x_j$ and $y_j$ are not ordered sequences, they are paired size-wise which again minimizes the denominator.

We now prove the general result by induction. The base case ($N=2$) is
\begin{equation}
\frac{(x_0+y_1)(x_1+y_0)}{(x_1+y_1)(x_0+y_0)} \ge 1,
\end{equation}
valid for both increasing and decreasing monotonic orderings of $x_j$ and $y_j$ by (\ref{reverse}). Assuming $R_N\ge1$ for $N$ transitions, the case $N+1$ satisfies
\begin{align}
R_{N+1}
&=\frac{(x_0+y_1)(x_1+y_2)\dots(x_{N-1}+y_N)(x_N+y_0)}
{(x_1+y_1)(x_2+y_2)\dots(x_N+y_N)(x_0+y_0)} \nonumber\\
&=R_N \frac{(x_{N-1}+y_N)(x_N+y_0)}{(x_N+y_N)(x_{N-1}+y_0)}. \label{fraction''}
\end{align}
If $y_N \ge y_0$, the transition $x_{N-1}\!\rightarrow\!x_N\!\rightarrow\!x_{N-1}$ is only possible for $x_N\ge x_{N-1}$; similarly for $y_N<y_0$, we require $x_N<x_{N-1}$. In both cases, the remaining fraction in (\ref{fraction''}) is $\ge1$ by (\ref{reverse}). Since $R_N\ge1$, it follows that $R_{N+1}\ge1$. Thus $\Delta \mathscr{I}_N \ge 0$ for any finite sequence.

Finally, let $\mu_N$ converge uniformly to $\mu(t)$ as $N\!\to\!\infty$. When $g$ satisfies the Lipschitz condition, the corresponding solutions of (\ref{diffeqn}) converge uniformly to $m$, and $\Delta\mathscr{I}_N$ converges to $\oint_C d\mathscr{I}$. Since each $\Delta\mathscr{I}_N \ge 0$, we have $\oint_C d\mathscr{I} \ge 0$.
\end{proof}

This result, though algebraically straightforward, is conceptually nontrivial. The inequality is not a generic property of all functions $H(\mu,m)$ but depends critically on the monotonic relations between variance, mean, and equilibrium sampling established earlier. Without these assumptions, the sign of the mixed derivative and the ordering arguments in the discrete proof would not hold. The theorem therefore captures a specific physical regularity: whenever variability increases with stimulus intensity, a sensory system that samples and adapts according to the rules must accumulate non-negative net information over any closed stimulation cycle. The deeper implications of this inequality—and the broader intepretation of sensory information—are discussed in the next section.

\section{Discussion}
A minimal and general mathematical description of sensory transduction—the ideal sensory unit—is proposed in which firing rate serves as a quantitative measure of uncertainty. The formulation reveals a striking formal correspondence with thermodynamic principles: Despite its nonlinear dynamics, the firing rate behaves as a state function, and sensory information, defined as the reduction of uncertainty through sampling, obeys a second-law-like inequality. The result generalizes a recently published neural theory that not only matches experimental data across modalities but also yielded a new, previously unobserved prediction.  Because experimental evidence cannot yet verify every component of that theory, the present formulation of the ideal sensory unit is intentionally less restrictive, retaining only those elements that correspond closely to experimental observation. Through this formulation, we show that any cyclic input necessarily produces a non-negative net gain in sensory information. The analogy to thermal physics is striking: despite describing very different domains, both systems express an underlying directionality.

The theorem implies a simple yet profound rule: any sensory system excited and then returned to rest accumulates non-negative information. The net reduction of uncertainty through sampling can never be negative. This occurs because higher stimulus levels are accompanied by higher variability—an intrinsic property of natural signals. For the same difference in sample size, uncertainty is reduced more when the signal is high than when it is low.  The analogy with thermal physics helps clarify this asymmetry. In thermodynamics, Clausius' theorem asserts that heat flows spontaneously from hot to cold, establishing a bias in one direction. In the sensory domain, a similar bias arises from the monotonic scaling of variance with mean: higher stimulation levels carry greater potential for uncertainty reduction. The sensory analogue of “heat flows from hot to cold” is therefore “more uncertainty in larger signals.” Both relations encode a fundamental asymmetry that governs the behaviour of complex systems.

Unlike thermodynamics, where $dS_{\text{prod}}$ cannot be calculated directly, the state-space representation of the ideal sensory unit allows explicit computation of entropy production, since the equations account for both the equilibrium and non-equilibrium dynamics. Nevertheless, the analogy has limits. In thermodynamics, $dS_{\text{prod}} \ge 0$ locally at every step, whereas in the sensory theorem the inequality holds only globally, i.e. over a closed-loop cycle. Within that cycle, local decreases in sensory information can occur, corresponding to phases of de-adaptation when a stimulus is lowered or turned off.  It is important however to emphasize that the analogy to thermodynamics is formal rather than physical.  Nevertheless, one influential school of thought views statistical physics itself as inference under incomplete information \cite{jaynes1957information}, a perspective that is somewhat similar to the sensory problem considered here.

While the present approach is related in spirit to frameworks such as efficient coding, predictive coding, and the free-energy principle, it differs sharply in scope. Those theories assume specific optimization goals or hierarchical architectures. By contrast, the sensory information theorem makes no assumptions about coding strategies or inference mechanisms; it relies only on measurable physiological variables and on how sampling variance changes with input. It sets a general constraint on sensory processing that stands independently of these other frameworks. Just as importantly, it generates refutable predictions that can be tested directly at the level of firing rate, unlike some alternative approaches.

To illustrate this point, consider how the sensory information theorem can be tested experimentally. We begin by recasting the theorem entirely in terms of firing rates rather than uncertainty. In the high/low stimulus paradigm, during the stimulus-on phase, the firing rate rises to a peak (PR) and, if the stimulus is held long enough, adapts to a lower steady-state (SS). After stimulus offset, the firing rate drops below baseline to a trough (TR) before gradually returning to the spontaneous rate (SR). To evaluate the sensory information inequality in this context, we calculate the sum of firing-rate changes across the full cycle, $(\text{PR} - \text{SS}) + (\text{TR} - \text{SR})$. If the adaptation response follows the predictions of Section~\ref{example}, then SS obeys (\ref{inequality}), and TR satisfies a similar relationship $\sqrt{\text{TR} \times \text{SS}} \le \text{SR} \le (\text{TR} + \text{SS})/2$, which can be rearranged to yield $\text{TR} \ge 2\text{SR} - \text{SS}$. Taking the upper bound for SS and the lower bound for TR, the full-cycle calculation gives $(\text{PR} - \text{SS}) + (\text{TR} - \text{SR}) \ge 0$, showing that in this limited example the theorem is obeyed. Thus, to verify (\ref{sensorysecondlaw}) experimentally, one only needs to demonstrate that the measured responses obey the relevant adaptation inequalities similar to what is observed in Fig.~1.

Further considerations arise for sensory information in the context of \textit{reversibility}. In classical thermodynamics, a reversible process is one in which both system and surroundings can, in principle, be restored to their original states, while an irreversible process leaves a permanent change in either the system or its surroundings. Entropy production vanishes only in the reversible limit, $dS_{\text{prod}}=0$. In the present framework, the lower bound in (\ref{sensorysecondlaw}) is achieved only when a stimulus is introduced so gradually that no excess $H_{\text{relax}}$ is generated—that is, when the system remains in equilibrium throughout. In the language of information theory, stimulus changes must be introduced with minimal “surprise,” allowing the system to react fully to each change before the next occurs. Practically, this means that the stimulus must vary on a timescale much slower than the intrinsic adaptation time.

As it turns out, irreversibility has very important consequences for perception. Sensory information, which is only non-zero when a stimulus is introduced irreversibly, has been used extensively to unify many aspects of the perceptual decision-making process \cite{norwich1989informational,norwich1993information,wong1997physics}. The hypothesis is that a stimulus change is only noticeable when sensory information $|\int_{m_0}^{m_1} d\mathscr{I}|$ exceeds a constant.  That is, a minimal amount of information must be gained or lost before a decision can be made. The formulation has been shown to account for a wide range of perceptual tasks—including absolute detection, difference thresholds, and reaction times—and unifies several well-known empirical laws governing perceptual decision-making including those of Weber, Pierson, and Bloch.  

Moreover, this same relation can be recast in terms of a temporal change in firing rate, $|\Delta F| = |F(I,0) - F(I,t)| \ge \text{constant}$, where $F(I,t)$ is calculated from (\ref{adaptation}) \cite{norwich1989informational}. This implies that a minimal change in firing rate is required before a stimulus can be perceived or acted upon. Although the full exploration of this hypothesis must await a future publication, similar “change-detection” rules have been widely employed in other sensory studies: a stimulus becomes perceptible only when evoked neural pattern is sufficiently distinct from a baseline to be recognized as being new (i.e. a ``change detector''). Sensory information represents the cumulative quantity required for decision-making.


%
Finally, the generality of the mathematical framework suggests that similar mechanisms may underlie not only sensory adaptation but also gain control, habituation, and even behavioural or affective processes. Hedonic adaptation, or the “hedonic treadmill,” provides an illustrative analogy (e.g. \cite{solomon1974opponent}). The emotional response to new or improved situation—such as increased utility or happiness—diminishes once novelty fades. A person who buys a new bicycle may initially feel joy and convenience, but these feelings wane over time. If the bicycle is later stolen, they can feel devastated; yet eventually, they adapt, finding new methods of transportation and returning to a baseline level of well-being. On the surface, nothing seems gained or lost. Yet a deeper principle suggests that, despite the eventual loss, the experience retains value. If the memory of the event persists, and given the net positive information gain theorem, one might say that despite both rise and fall, the system is left richer—a poetic echo of Tennyson’s reflection that `Tis better to have loved and lost than never to have loved at all.’

\appendix
\section*{Appendix A: Variables and Parameters}

Key symbols and their meanings are summarized below.  

\begin{table}[h!]
\caption{Principal symbols and their meanings.}
\label{tab:variables}
\centering
\scriptsize
\begin{tabular}{ll}
\hline
\textbf{Symbol} & \textbf{Meaning} \\
\hline
\\
\multicolumn{2}{l}{\textbf{(a) Ideal sensory unit}}\\
$\mu(t)$        & Input stimulus intensity \\
$m(t)$          & Sample size (state variable) \\
$m_{eq}(\mu)$   & Equilibrium sample size for given $\mu$ \\
$g(m,m_{eq})$   & State dynamics; $\dot m = g(m,m_{eq})$ \\
$F(t)$          & Firing rate (observable output) \\
$H$             & Shannon uncertainty, $\tfrac12\log(\sigma_R^2+\sigma^2/m)+c$ \\
$\sigma_R^2$    & Constant representation noise \\
$\sigma^2(\mu)$ & Stimulus variance (typically $\propto \mu^p$) \\
$k$             & Proportionality constant in $F=kH$ \\[4pt]
\\
\multicolumn{2}{l}{\textbf{(b) Example model (Section~\ref{example})}}\\
$a$             & Relaxation rate in $\dot m=-a(m-m_{eq})$ \\
$I$             & Mean stimulus level; $\mu = I + \delta I$ \\
$\delta I$      & Additive stimulus noise \\
$p$             & Fluctuation (Tweedie) exponent \\
$\beta$         & Scale constant for dimensional consistency \\
SR              & Spontaneous firing rate \\
PR              & Peak firing rate \\
SS              & Steady-state firing rate \\
TR              & Trough firing rate after offset \\[4pt]
\\
\multicolumn{2}{l}{\textbf{(c) Analogy with thermodynamics}}\\
$H_{eq}$        & Uncertainty at equilibrium (state function) \\
$F_{eq}$        & Firing rate at equilibrium (state function) \\
$\delta H_{\text{flux}}$  & $(\partial H/\partial\mu)\,d\mu$ (flux term) \\
$\delta H_{\text{relax}}$ & $(\partial H/\partial m)\,dm$ (relaxation term) \\
$\mathscr{I}$              & Sensory information: $-\delta H_{\text{relax}}$ \\
$S$              & Thermodynamic entropy \\
$S_{\text{prod}}$ & Entropy production \\
$Q$              & Heat exchanged \\
$T$              & Temperature (state variable) \\
$P, V$          & Pressure and volume (state variables) \\
\hline
\end{tabular}
\end{table}

\bibliographystyle{spphys}
\bibliography{template}   


\end{document}